\documentclass[aps,twocolumn]{revtex4}
\usepackage{graphicx}

\begin{document}
\title{Two-terminal transport in biased lattices: transition from ballistic to diffusive current}
\author{Andrey R. Kolovsky}
\affiliation{Kirensky Institute of Physics, 660036, Krasnoyarsk, Russia} 
\affiliation{Siberian Federal University, 660041 Krasnoyarsk, Russia} 
\date{\today}

\begin{abstract}
We analyze quantum transport of charged fermionic particles in the tight-binding lattice connecting two particle reservoirs (the leads). If the lead chemical potentials are different they create an electric field which tilts the lattice. We study the effect of this tilt on quantum transport in the presence of weak relaxation/decoherence processes in the lattice. It is shown that the Landauer ballistic transport regime for a weak tilt (small chemical potential difference) changes to the diffusive Esaki-Tsu transport regime for a strong tilt (large chemical potential difference), where the critical tilt for this crossover is determined by the condition that the Wannier-Stark localization length coincides with the lattice length.
\end{abstract}
\maketitle

%
{\em 1.} Theory of quantum transport in biased lattices is traced back to the work by Zener \cite{Zene34} who noticed that crystalline electrons subject to an electric field should show periodic oscillation (known nowadays as Bloch oscillation) but not directed current which was typically observed in laboratory experiments. This contradiction between theory and experiment was resolved by realizing that directed current in solid crystals is due to an interplay between Bloch oscillation and inelastic scattering of electrons on phonons. Denoting the scattering rate by $\gamma$ and the Bloch frequency by $F$ (which in dimensionless units used through the paper coincides with the electric field magnitude) this current obeys the celebrated Esaki-Tsu equation \cite{Esak70}
\begin{equation}
\label{esaki} 
\bar{j} \sim \frac{F/\gamma}{1+(F/\gamma)^2} \;,
\end{equation}
which interpolates between the liner response regime for $F\ll\gamma$ and negative differential conductivity regime for $F\gg\gamma$. We mention that Eq.~(\ref{esaki}), which we shall discuss in more detail later on, implies an infinite lattice.
\begin{figure}[b]
\includegraphics[width=7.5cm,clip]{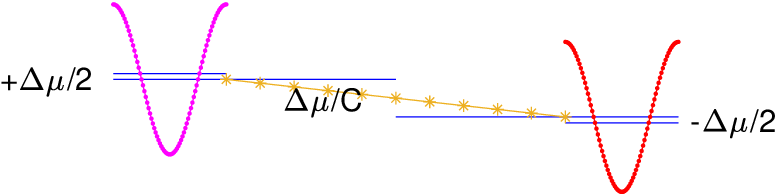}
\caption{Pictorial presentation of the model where a finite size tight-binding chain (asterisks) connects two reservoirs of fermionic particles with the energy spectrum given by the cosine dispersion relation.}
\label{fig1}
\end{figure}

One finds a completely different formulation of the transport problem for finite lattices. Here the unbiased lattice of the length $L$ is attached to particle reservoirs (the leads). If the lead chemical potentials are different there is a directed current across the lattice. The common theoretical approach to this transport problem is the Landauer theory \cite{Land57, Buet85,Datt95} which relates the system conductance to the transmission probability for the Bloch wave with the Fermi energy. It should be stressed that, unlike Eq.~(\ref{esaki}) which describes the diffusive current, the stationary current in the considered two-terminal setup is ballistic. Let us also mention that there is no lattice tilt in the standard Landauer approach.

To this end we introduce the model which unifies the above transport problems. We shall model the leads by tight-binding rings where, to be certain, we assume half-filling of ring sites. It is also assumed that this equilibrium situation corresponds to the vanishing electrostatic energy of the leads. At time $t=0$ we change the lead chemical potentials by a small but finite value $\pm \Delta\mu/2$. This changes the lead electrostatic energies by $\Delta E=\pm \Delta\mu/2C$ where $C$ is the lead capacity. Simultaneously, this tilts the lattice connecting the leads, see Fig.~\ref{fig1}. We are interested in the current across the lattice. This problem was addressed in the recent work \cite{Pinh23} where the authors focused on transient dynamics by using the Green function formalism. Here we use the other approach of the master equation for carrier's density matrix and focus exclusively on the steady-state regime \cite{remark1}. The main advantage of the master equation approach is that it allows us (i) to take into account a finite thermalization rate for carriers in the leads and (ii) to include into consideration relaxation processes in the lattice. Below we show how the Landauer ballistic transport regime transforms into the diffusive Esaki-Tsu regime depending on the system parameters.

%
{\em 2.} For future purposes we recall derivation of the Esaki-Tsu formula (\ref{esaki}). In the presence of relaxation processes dynamics of non-interacting carriers in a biased lattice is described by the master equation for the carrier single-particle density matrix $\rho=\rho(t)$,
\begin{equation}
\label{master_spdm} 
\frac{d \rho}{dt}=-i[H,\rho] - \gamma {\cal L}(\rho) \;, 
\end{equation}
where $H$ is the tight-binding Hamiltonian in the Wannier basis,
\begin{equation}
\label{tilted_spdm} 
H=-\frac{J}{2}\sum_{\ell} \left( |\ell+1\rangle \langle \ell | +h.c.\right)+ F\sum_{\ell} |\ell \rangle \ell\langle \ell | \;,
\end{equation}
and ${\cal L}(\rho)$ is the Lindblad relaxation operator \cite{Lind76}
\begin{eqnarray}
\label{lind1} 
{\cal L}(\hat{\rho})=\sum_{s,q} W(s,q) \left[\hat{\rho}\hat{\sigma}^{\dagger (s,q)}\hat{\sigma}^{(s,q)} 
\right. \\
\nonumber
\left.
- 2\hat{\sigma}^{(s,q)}\hat{\rho}\hat{\sigma}^{\dagger (s,q)} + \hat{\sigma}^{\dagger (s,q)}\hat{\sigma}^{(s,q)} \hat{\rho}\right] \;.
\end{eqnarray}
In this equation the operators $\hat{\sigma}^{(s,q)}$ are given in the quasimomentum basis by the matrices
$\sigma^{(s,q)}_{k,p}=\delta_{k,s}\delta_{p,q}$
%
and coefficients $W(s,q)$ have a sense of transition rates between different quasimomentum states. Playing with transition rates $W(s,q)$ one can address different relaxation dynamics \cite{77}. We shall consider the simplest case where the Lindblad operator Eq.~(\ref{lind1}) reduces to the following form 
\begin{equation}
\label{lind3} 
{\cal L}(\rho)=\rho - \rho_0 \;, \quad \rho_0=\sum_{|k|<k_F} | k\rangle\langle k | \;.
\end{equation}
In this case the stationary density matrix can be found analytically. Indeed, switching to the basis of the Wannier-Stark states
\begin{equation}
\label{WS} 
| n \rangle=\sum_l {\cal J}_{l-n}\left(\frac{J}{F}\right) | l \rangle 
\end{equation}
(here ${\cal J}_{m}(z)$ is the Bessel function of the first kind), the stationary solution of the master equation (\ref{master_spdm}) with the relaxation term (\ref{lind3}) reads \cite{Mino04}
\begin{equation}
\label{steady} 
\bar{\rho}_{n,n'}=\frac{\gamma}{\gamma-iF(n-n')} \langle n | \rho_0 | n' \rangle \;. 
\end{equation}
Then it is a matter of few lines to show that Eq.~(\ref{steady}) leads to the Esaki-Tsu formula for the stationary current \cite{95}.

%
{\em 3.} We come back to the model shown in Fig.~\ref{fig1}. First, we formalize this model and we begin with the case of no relaxation processes in the lattice. In this case the governing master equation has the standard form Eq.~(\ref{master_spdm}) where, however, the Lindblad relaxation operator (\ref{lind3}) acts only on the leads whose Hamiltonians $H_i$ have the form 
\begin{equation}
\label{rings} 
H_i=\pm \frac{\Delta\mu}{2C} - J\sum_k \cos\left(\frac{2\pi k}{M}\right) | k \rangle\langle k | \;.
\end{equation}
In Eq.~(\ref{rings}) the index $i=L,R$ labels the leads and the plus and minus sigh refers to the left and right rings which are modeled by the tight-binding rings of the size $M$. To simplify notations we omit index $i$ for the quasimomentum states but it should be remembered that they are different for the left and right leads. The Fermi quasimomenta $k_F$ in Eq.~(\ref{lind3}) are also different because the leads have different chemical potentials $\mu_L=\Delta\mu/2$ and $\mu_R=-\Delta\mu/2$. The Hamiltonian of the whole system is 
\begin{equation}
\label{total}
H_{tot}=H + \sum_{i=L,R}\left(H_i +V_i\right) \;,
\end{equation}
where the lattice Hamiltonian $H$ is given in Eq.~(\ref{tilted_spdm}) with the site index $\ell$ running from $\ell=1$ to $\ell=L$. Finally, the coupling Hamiltonians are
\begin{eqnarray}
\label{epsilon}
V_L= -\frac{\epsilon}{2} \frac{1}{\sqrt{M}}\sum_{k=1}^M | k \rangle\langle \ell=1 | \;, \\
\nonumber
V_R=-\frac{\epsilon}{2} \frac{1}{\sqrt{M}}\sum_{k=1}^M | k \rangle\langle \ell=L | \;,
\end{eqnarray}
where we introduced the coupling constant $\epsilon\ll J$. 

To find the stationary current between the leads we use numerical approach of Ref.~\cite{131}. In brief, we set the left-hand-side of the master equation to zero and solve the obtained algebraic equation for the steady-state total density matrix $\bar{\rho}_{tot}$ which has dimension $(M+L+M)\times(M+L+M)$. Our interest, however, is not the total density matrix but the lattice density matrix $\bar{\rho}$ which has dimension $L\times L$. We stress that we always check convergence of this matrix with respect to the limit $M\rightarrow\infty$ which relates the master equation approach to the Landauer theory. In fact, as it was demonstrated in Ref.~\cite{131}, for unbiased lattices ($C=\infty$) the stationary solution of the considered master equation reproduces the Landauer result in the limit $M\rightarrow\infty$ and $\gamma\rightarrow 0$ where the former limit should be taken first. 
\begin{figure}
\includegraphics[width=8.5cm,clip]{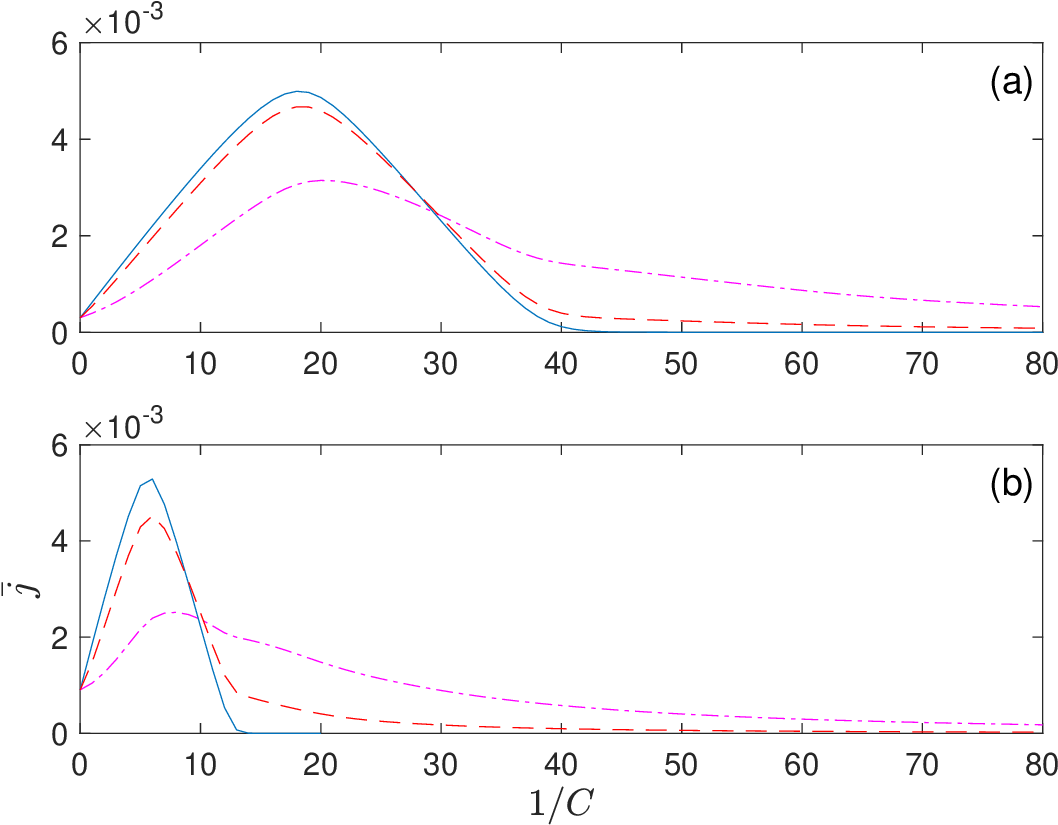}
\caption{Stationary current across the tilted lattice of the length $L=40$ (a) and $L=120$ (b). The system parameters are $J=1$, $\epsilon=0.1$, $\gamma=0.2$, and $\Delta\mu =\pi/60 \approx 0.05$ (a) and $\Delta\mu =\pi/20$ (b). The blue solid lines, red dashed lines, and magenta dash-dotted lines correspond to $\tilde{\gamma}=0$, $\tilde{\gamma}=2\cdot 10^{-5}$, and $\tilde{\gamma}=2\cdot 10^{-4}$, respectively.}
\label{fig2}
\end{figure}

%
{\em 4.} The solid blue lines in Fig.~\ref{fig2} shows the stationary current across the lattice of the length $L=40$ (upper panel) and $L=120$ (lower panel) as the function of the lead capacity $C$ for a small but finite $\Delta\mu \ll J$ and $\gamma=0.2$. Notice that the lead capacity determines the electric field magnitude through the equation
\begin{equation}
\label{field}
F=\frac{\Delta\mu}{CL} \;.
\end{equation}
Notice that we use different $\Delta\mu$ for different $L$ to keep $F$ the same. It is seen in Fig.~\ref{fig2} that a weak electric field (small values of the control parameter $1/C$) enhances the stationary current but strong electric field suppresses it. This suppression is obviously due to the Wannier-Stark localization where one expects the current to vanish completely as soon as the Wannier-Stark localization length $L_{WS}\approx 2J/F$ becomes smaller than the lattice length $L$. For parameters of Fig.~\ref{fig2}(a) the critical point where $L_{WS}=L$ corresponds to $1/C\approx 40$ ($F_{cr}=0.05$) and it is three times smaller for parameters of Fig.~\ref{fig2}(b). 

\begin{figure}
\includegraphics[width=8.0cm,clip]{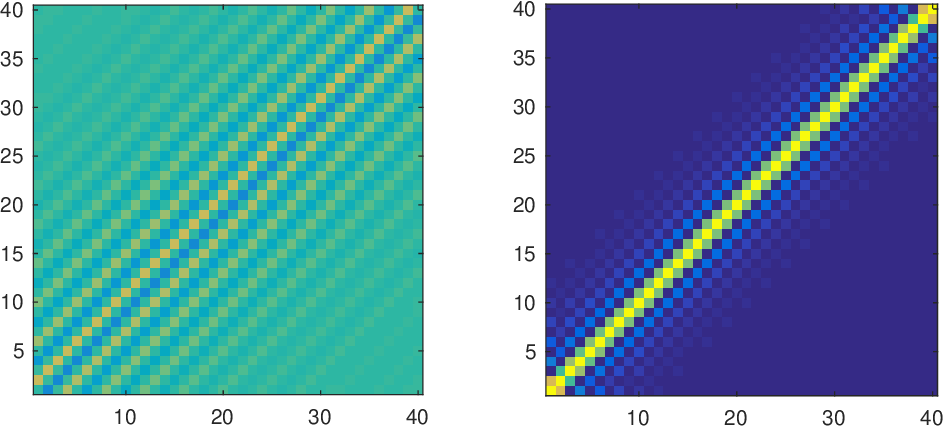}
\caption{The imaginary part of the matrix elements $\bar{\rho}_{\ell,\ell'}$ (left panel) and their absolute values (right panel) for $1/C=0$. The other parameters are $J=1$, $\epsilon=0.1$, $\gamma=0.2$, and $\Delta\mu=\pi/20$.}
\label{fig4a}
\end{figure}

\begin{figure}
\includegraphics[width=8.0cm,clip]{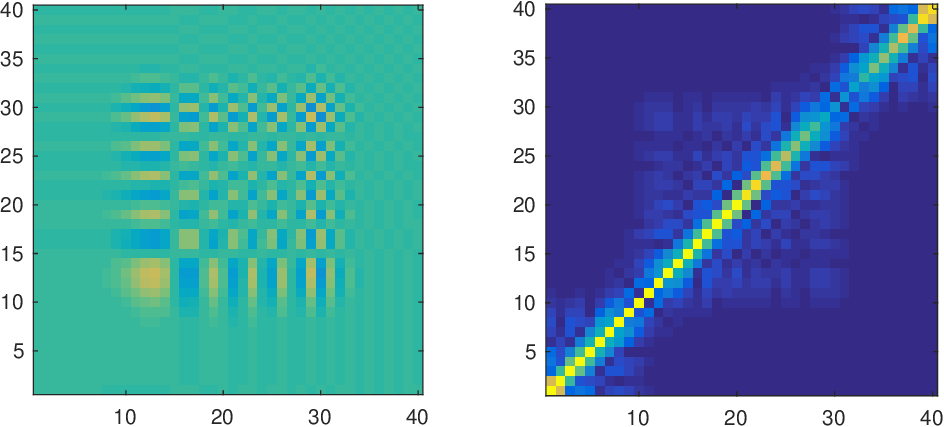}
\caption{The same as in Fig.~\ref{fig4a} yet for $1/C=25$ where the Wannier-Stark localization length is smaller then the lattice length.}
\label{fig4b}
\end{figure}

One gets a deeper insight into the discussed Wannier-Stark localization by analyzing the structure of the steady-state density matrix $\bar{\rho}$. As two examples, Fig.~\ref{fig4a} and Fig.~\ref{fig4b} show the imaginary part of the matrix elements $\rho_{\ell,\ell'}$ (left panels) and their absolute values (right panels) for $F=0$ and $F>F_{cr}$, respectively. While for $F=0$ the matrix $\bar{\rho}$ is a band matrix, it evolves in the interval $0<F<F_{cr}$ into a matrix where its central part is associated with the localized Wannier-Stark states. 

Till now we have considered cases where relaxation processes take place only in the leads. Remarkably, this suffices for the whole system to relax into a steady state. In the other words, Bloch oscillations induced by a sudden tilt of the lattice decay in course of time and the current takes a constant value depicted by blue solids lines in Fig.~\ref{fig2}. We mention that this steady-state current is ballistic in the sense that, similar to the case $F=0$, it depends only on the reservoir chemical potential difference $\Delta\mu$ but not on the lattice length $L$. The reason behind this universality is that at fixed $\Delta\mu$ and increasing $L$ the Wannier-Stark localization length increases proportionally to $L$. The intrinsic relation of the currently discussed ballistic transport regime to the Landauer theory also follows from the initial linear growth of the current, $\bar{j}\sim \Delta\mu +\Delta\mu/C$. In fact, this linear dependence immediately follows from the Landauer formula,
\begin{equation}
\label{landauer}
\bar{j} \sim \int_{-\frac{\Delta\mu}{2}-\frac{\Delta\mu}{2C}}^{+\frac{\Delta\mu}{2}+\frac{\Delta\mu}{2C}} |t(E)|^2 dE \;,
\end{equation}
if we neglect the dependence of the transmission probability $|t(E)|^2$ on the energy. The detailed comparison of the depicted in Fig.~\ref{fig2} numerical results with Eq.~(\ref{landauer}) is given in Ref.~\cite{134}. 
\begin{figure}
\includegraphics[width=8.0cm,clip]{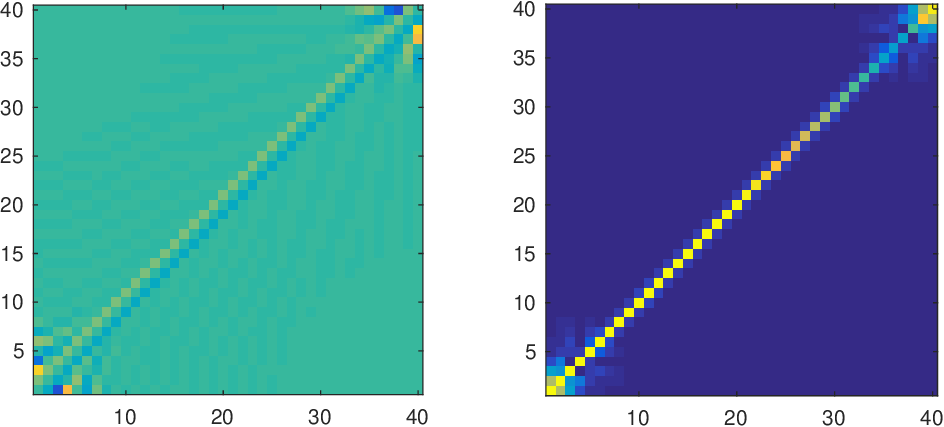}
\caption{The same as in Fig.~\ref{fig4b} yet for $1/C=80$ and finite decoherence rate $\tilde{\gamma}=2\cdot1 0^{-4}$.}
\label{fig4c}
\end{figure}

%
{\em 5.} Next we address the diffusive current, in the spirit of the Esaki-Tsu theory. To do this we include in the master equation the Lindblad relaxation operator which acts on carriers in the lattice. Namely, we shall consider the following relaxation term,
\begin{equation}
\label{decoherence}
\widetilde{{\cal L}}(\rho)_{\ell,\ell'}= -\tilde{\gamma}(1-\delta_{\ell,\ell'})\rho_{\ell,\ell'} \;,
\end{equation}
which causes decay of off-diagonal elements of the lattice density matrix with the rate $\tilde{\gamma}$. We mention that for infinite lattice the master equation with the relaxation term Eq.~(\ref{decoherence}) describes the decaying Bloch oscillations with vanishing steady-state current \cite{77}. This is, however, not the case of a finite lattice attached to the leads. The dashed and dash-dotted lines in Fig.~\ref{fig2} show the stationary current between the leads for $\tilde{\gamma}=2\cdot 10^{-5}$ and $\tilde{\gamma}=2\cdot 10^{-4}$, respectively. It is seen that a weak decoherence process in the lattice destroys the Wannier-Stark localization and originates a non-zero stationary current. Numerical analysis of this current reveals the following functional dependence
\begin{equation}
\label{diffusive}
\bar{j}\sim \frac{\tilde{\gamma} J}{F^2 L} = \frac{\tilde{\gamma} L J}{\Delta\mu^2} \left(\frac{1}{C}\right)^{-2} \;,
\end{equation}
which can be referred to as the negative differential conductivity regime for finite lattices. For the sake of comparison with the case $\tilde{\gamma}=0$ Fig.~\ref{fig4c} shows the carrier density matrix in this regime. It is seen that far from the lattice edges this is a three-diagonal matrix. In the rest of the work we derive Eq.~(\ref{diffusive}) by using a semi-analytical approach.

\begin{figure}
\includegraphics[width=8.5cm,clip]{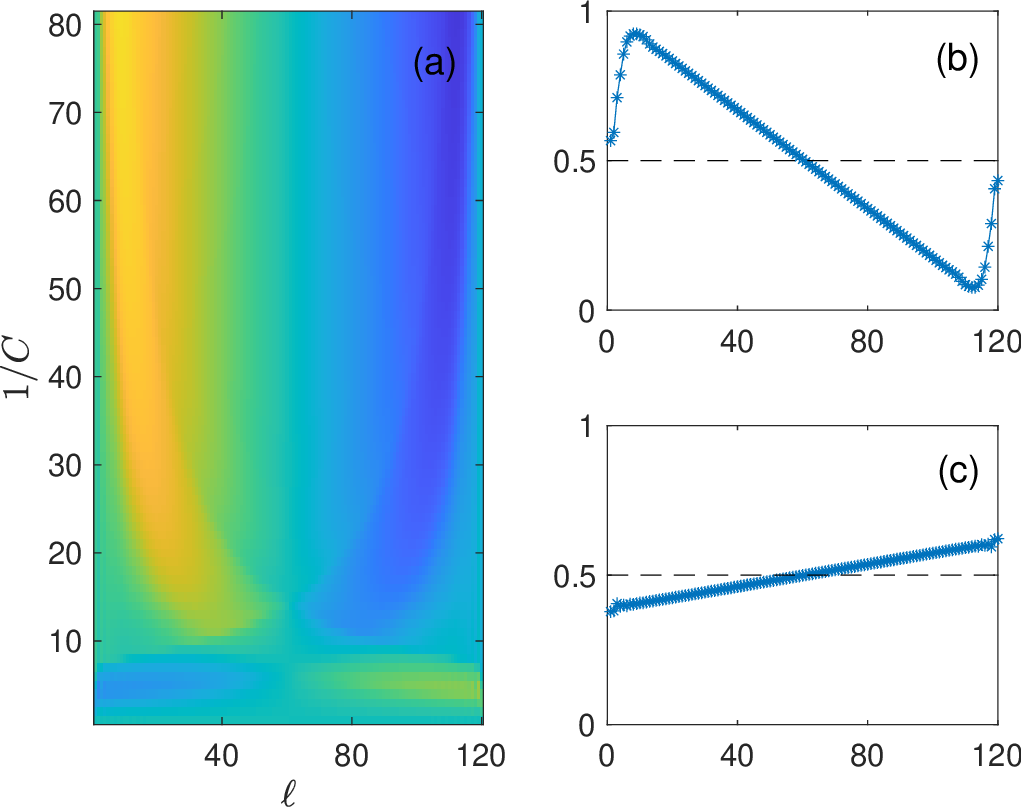}
\caption{(a) Populations of the lattice sites as a color map. The panel (b) an (c) show populations for $1/C=80$ and $1/C=3$, respectively. The dashed line corresponds to the equilibrium situation ($\Delta\mu=0$) where $\bar{\rho}_{\ell,\ell}=0.5$. The system parameters are $L=120$, $\gamma=0.2$, and $\tilde{\gamma}=2\cdot10^{-6}$.}
\label{figA1}
\end{figure}

%
{\em 6.} First, we discuss non-equilibrium population of the lattice sites, -- the result which we shall use later on to obtain the estimate Eq.~(\ref{diffusive}). It is seen in Fig.~\ref{fig2} that a weak decoherence strongly affects the steady-state current across the lattice. Since the current is proportional to the imaginary part of the off-diagonal matrix elements $\bar{\rho}_{\ell,\ell+1}$ this means that non-zero $\tilde{\gamma}$ strongly affect off-diagonal matrix elements. However, due to the specific structure of the relaxation operator Eq.~(\ref{decoherence}) it has a minor effect on diagonal elements $\bar{\rho}_{\ell,\ell}$ which appear to be mainly determined by the field magnitude $F\sim 1/C$.  The lattice site populations as the function of the parameter $1/C$ are shown in Fig~\ref{figA1}. Interestingly, in the Landauer regime site populations increase with an increase of the site index $\ell$ while for $F>F_{cr}$ the situation is inverted.

Now we are prepared to derive the estimate Eq.~(\ref{diffusive}) for the diffusive current. Let us focus on the central part of the lattice where the ordinary differential equation for the matrix elements $\rho_{\ell,\ell+1}$ reads
\begin{eqnarray}
\label{a1}
\dot{\rho}_{\ell,\ell+1}=-i F\rho_{\ell,\ell+1} - i\frac{J}{2} \left(\rho_{\ell,\ell+2} + \rho_{\ell,\ell} \right. \\
\nonumber
\left. -\rho_{\ell+1,\ell+1} -\rho_{\ell-1,\ell+1} \right) - \tilde{\gamma}\rho_{\ell,\ell+1} \;.
\end{eqnarray}
Next, we set the left-hand-side of this equation to zero and, basing on the numerical result shown in Fig.~\ref{fig4c}, neglect the elements $\rho_{\ell,\ell+2}$ and $\rho_{\ell-1,\ell+1}$. Denoting by $x$ and $y$ the real and imaginary parts of $\rho_{\ell,\ell+1}$ we have
\begin{equation}
\label{a2}
-i F(x+iy) -i\frac{J}{2}\frac{d\rho_{\ell,\ell}}{d\ell} -\tilde{\gamma}(x+iy) = 0 \;.
\end{equation}
\vspace{0.0cm}
For a small $\tilde{\gamma}$ solution of this equation is
\begin{equation}
\label{a3}
y\approx \frac{\tilde{\gamma}J}{2F^2} \frac{d\rho_{\ell,\ell}}{d\ell} \;.
\end{equation}
Finally, considering the limit of large $F$ we can approximate the derivative $d\rho_{\ell,\ell}/d\ell$ by $1/L$, that gives the estimate (\ref{diffusive}).

%
{\em 7.} To summarize, we analyzed quantum transport of non-interacting spinless fermions in the tilted lattice which connects two reservoirs of Fermi particles with different chemical potentials. As it was already mentioned, this problem was addressed earlier in Ref.~\cite{Pinh23} by using Green functions formalism. We revisited this problem by using the master equation approach. We confirmed the main result of the cited work that the stationary current between the leads vanishes if the lattice length exceeds the Wannier-Stark localization length. This condition defines the critical field magnitude $F_{cr}$). We notice that, unlike in Ref.~\cite{Pinh23}, we do not consider the filed magnitude $F$ as an independent parameter but relate it to the chemical potentials of the leads and their capacities.

The new results are due to making account of relaxation/decoherence processes in the system. It was found that a weak decoherence destroys the Wannier-Stark localization and originates the diffusive current in the parameter region $F>F_{cr}$. (By `weak decoherence' we mean that it has a minor effect on the ballistic current in the parameter region $F<F_{cr}$.) This transition from ballistic to diffusive currents was well observed in our numerical experiments. Since a weak decoherence is inevitably present in any laboratory experiment the reported results can guide the searching of this transition in real physical systems.



\begin{thebibliography}{99}
\bibitem{remark1}
 Results on transient dynamics of the system will be published elsewhere \cite{134}.
\end{thebibliography}
\end{document}